\documentclass[10pt,conference]{IEEEtran} 
\IEEEoverridecommandlockouts
\usepackage{graphicx}
\usepackage{amsmath}
\usepackage{amssymb}
\usepackage{algorithm}
\usepackage{algorithmic}
\usepackage[T1]{fontenc}
\usepackage{color}
\usepackage{subfigure}
\usepackage{booktabs}
\usepackage{cases}
\usepackage{cite}
\usepackage{footmisc}
\usepackage{CJK}
\usepackage{type1cm}
\usepackage{times}
\usepackage{multicol}
\usepackage{multirow}
\usepackage{setspace}
\usepackage{array}
\usepackage{adjustbox}
\usepackage{tabularx}
\usepackage{booktabs}
\usepackage{overpic}

\def\BibTeX{{\rm B\kern-.05em{\sc i\kern-.025em b}\kern-.08em
    T\kern-.1667em\lower.7ex\hbox{E}\kern-.125emX}}

\usepackage{fancyhdr}
\fancypagestyle{firststyle}{\fancyhf{}
	\fancyhead[L]{\small M. Ying, D. Shakya, and T. S. Rappaport, ``Using Waste Factor to Optimize Energy Efficiency in Multiple-Input Single-Output (MISO) and Multiple-Input Multiple-Output (MIMO) Systems", in\textit{ 2024 IEEE Global Communications Conference (GLOBECOM), } Cape Town, South Africa, Dec. 2024, pp. 1--6.}
}

\usepackage{times}
\begin{document}
	
\title{Using Waste Factor to Optimize Energy Efficiency in Multiple-Input Single-Output (MISO) and Multiple-Input Multiple-Output (MIMO) Systems}
% Extending Waste Factor Analysis: From SISO to MIMO System Power Efficiencies

\author{
	\IEEEauthorblockN{Mingjun Ying, Dipankar Shakya, and Theodore S. Rappaport}
\IEEEauthorblockA{ NYU WIRELESS, Tandon School of Engineering, New York University, Brooklyn, NY, 11201\\
 \{yingmingjun, dshakya, tsr\}@nyu.edu}
 \thanks{This research is supported by the NYU WIRELESS Industrial Affiliates Program and the NYU Tandon School of Engineering graduate fellowship.}
 \vspace{-15pt}
 }

\maketitle
% \linespread{0.95}
\newcommand\blfootnote[1]{%
	\begingroup
	\renewcommand\thefootnote{}\footnote{#1}%
	\addtocounter{footnote}{-1}%
	\endgroup
}

	% \blfootnote{\noindent{This research is supported by the New York University (NYU) WIRELESS
	% Industrial Affiliates Program, and NYU Tandon School of Engineering graduate fellowship.}}

\thispagestyle{firststyle}
\begin{abstract}

This paper introduces Waste Factor (\textit{W}) and Waste Figure (\textit{WF}) to assess power efficiency in any multiple-input multiple-output (MIMO) or single-input multiple-output (SIMO) or multiple-input single-output (MISO) cascaded communication system. This paper builds upon the new theory of Waste Factor, a systematic model for added wasted power in any cascade for parallel systems such as MISO, SIMO, and MIMO, which are prevalent in current wireless networks. Here, we also show the advantage of \textit{W} compared to conventional metrics for quantifying and analyzing energy efficiency. This work explores the utility of \textit{W} in assessing energy efficiency in communication channels, within Radio Access Networks (RANs).

\end{abstract}
	

\section{Introduction}

This paper presents new results for using Waste Factor (\textit{W}) or Waste Figure (\textit{WF}) in dB for power efficiency of parallel circuits and system configurations such as found in MISO, SIMO, and MIMO wireless systems.  

$W$ has shown promise for quantifying and optimizing energy efficiency in diverse applications, including UAV communication systems \cite{xing2021high}, massively broadband systems \cite{Murdock2011GC}, millimeter wave wireless networks, phase shifters in sub-THz phased arrays \cite{kanhere2022power}, wireless relay networks \cite{murdock2014JSAC}, data centers \cite{ying2023waste, Rappaport2024Microwave}, and circuits \cite{Rappaport2024Microwave}, yet its broader adoption has been somewhat limited due to its very recent development and also because the application of \textit{W} in parallel systems has not been previously explored, or defined.

Inspired by Friis's 1944 analysis of additive noise in cascaded systems \cite{friis1944}, we adopt a similar theory and model for additive wasted power, thus providing \textit{W} as a unified Key Performance Indicator (KPI) to evaluate the energy efficiency of any cascaded or paralleled system. Then, we show the application of $W$ in diverse system configurations such as MISO, SIMO, and MIMO.

\section{Advantages of the Waste Factor Compared to Conventional Energy Efficiency Metrics}

There is a tremendous focus on energy efficiency in wireless networks.

Previously used energy metrics include the ratio between the total number of packets received at the destination node and the total energy consumption spent by the network to deliver these packets \cite{boyle2017energy}, and defined for Internet of things (IoT) as the ratio between `How many messages are received' and `How many messages could have been received regarding the total energy consumption' \cite{hossfeld2022analytical}.

In \cite{Mccune2019globecom}, McCune highlights that energy efficiency in wireless links varies by more than eleven orders of magnitude and stresses the need for a uniform approach to energy efficiency across different systems. In \cite{chen2010energy}, the authors study the component, equipment, and system/network levels, and find energy efficiency metrics at the component and equipment levels are well-developed, while those for system/network levels need more attention. 

In \cite{murdock2014JSAC}, the consumption factor (precursor to Waste Factor) evaluates the energy efficiency of a relay network and cascaded circuits. In \cite{kanhere2022power}, the consumption efficiency factor (CEF) provides a quantitative metric for the trade-off between the data rate and the power consumed by a communication system using Waste Factor and provides insights for network energy efficiency with different cell sizes.

We first motivate the use of Waste Factor theory by highlighting several key advantages of $W$ over previous energy efficiency metrics:
\begin{itemize}
    \item \textbf{Comprehensive Analysis}: \(W\) provides a unified assessment of power efficiency, covering every aspect of the system power usage, including signal, non-signal, and ancillary power consumption \cite{ying2023waste, Rappaport2024Microwave, Rappaport2024Access, Rappaport2024WCNCTutorial}.

    \item \textbf{System-Wide Applicability}: Unlike conventional metrics focused on specific components or levels, \(W\) is applicable to 
    any source-to-sink communications architecture, including SIMO, MISO and MIMO as shown here.

    \item \textbf{Optimization and Strategic Design}: \(W\) not only offers precise insights into the sources of power waste within a system but also facilitates targeted enhancements, enabling comparative analyses and informed decision-making in early design stages \cite{Rappaport2024Microwave, ying2023waste, Rappaport2024Access, Rappaport2024WCNCTutorial}. 
\end{itemize}

\section{Noise Factor and Waste Factor Fundamentals}

\subsection{Quantifying Additive Noise with Noise Factor}

Noise Factor (\textit{F}) quantifies the additive noise contributed by each component in a cascade, and is crucial for analyzing signal-to-noise ratio (SNR) degradation in a cascade, where $F = \text{SNR}_{i}/\text{SNR}_{o}$~\cite{friis1944}. 

The total noise factor for a cascade with matched loads, from the source to sink, was derived by Friis to be \cite{friis1944}:
\begin{equation}
\label{Fcas}
    F = F_1 + \frac{(F_2 - 1)}{G_1} + \frac{(F_3 - 1)}{G_1 G_2} + \ldots + \frac{(F_N - 1)}{\prod_{i=1}^{N-1} G_i},
\end{equation}
where \( F_i \) and \( G_i \) represent the noise factor and power gain of the \(i^{th}\) device, respectively. The additional noise power contributed by a device in the cascade (independent of the input noise power) is quantified by:
\begin{equation}
\label{addnoise}
    P_{\text{additive-noise}} = (F_{i} - 1)G_{i}N_{i},
\end{equation}
where $N_i$ is the noise power of \(i^{th}\) device input.

\subsection{Quantifying Additive Wasted Power with Waste Factor}
Waste Factor is derived using an approach similar to Friis's Noise Factor, yet is applied to systematically quantify the wasted power in a cascade. $W$ can be used in all types of systems that involve the transmission of information (e.g. which have a source and a sink), including circuits, wireless links, propagation channels, and data centers \cite{Rappaport2024Microwave, ying2023waste, Rappaport2024Access, murdock2014JSAC, Murdock2011GC, kanhere2022power}. 

The total power consumed by any system or cascade is directly related to the conservation of power in these four power components \cite{murdock2014JSAC, Murdock2011GC, Rappaport2024Microwave}:  
1)\textbf{\(P_{\text{source,out}}\)} is the power from a source that is applied at the input of the cascaded system; 2) \textbf{\(P_{\text{signal}}\)} is the signal power delivered to the output of the cascade or device; 3) \textbf{\(P_{\text{non-signal}}\)} is the power consumed by devices within the cascade (e.g., devices along the signal path) but which is not part of the delivered signal power and is viewed as wasted power which includes power consumed by passive or active devices or lost due to heat dissipation; 4)\textbf{\(P_{\text{non-path}}\)} is the power consumed by devices that are not on the signal path and do not contribute to the output signal power. However, as shown in \cite{Rappaport2024Access} and \cite{Rappaport2024Microwave}, quiescent power consumed by amplifiers on the signal path (when on stand-by) may be considered in $P_{\text{non-path}}$.

Waste Factor is defined in eq. \eqref{defW} as the ratio of the total power consumed by the signal path components along a cascade $P_{\text{consumed,path}}$, which is the sum of additive wasted power and the useful signal output power, to the useful signal output power \cite{Rappaport2024Microwave, ying2023waste, Murdock2011GC, Rappaport2024Access}, and $\eta_{\text{w}}$ is waste factor efficiency, which is defined as the reciprocal of $W$:
\begin{equation}
\vspace{-3pt}
W = \frac{1}{\eta_{\text{w}}}=\frac{P_{\text{consumed,path}}}{P_{\text{out}}} = \frac{P_{\text{signal}} + P_{\text{non-signal}} }{P_{\text{signal}}}.
\label{defW}
\end{equation}

The total power consumption of any system $P_{\text{consumed,total}}$ is the sum of power consumed on the signal path $P_{\text{consumed,path}}$ as well as all power consumed by non-signal path $P_{\text{non-path}}$, where $W$ is used to characterize power wasted for each component and the entire cascade on the signal path between source and sink \cite{Rappaport2024Microwave, ying2023waste, Murdock2011GC, murdock2014JSAC, Rappaport2024Access, Rappaport2024WCNCTutorial}:
\vspace{-3pt}
\begin{align}
\label{powercontotal}
   \nonumber P_{\text{consumed,total}} &= P_{\text{source,out}} + P_{\text{system,added}}\\& \nonumber+ P_{\text{non-signal}} + P_{\text{non-path}}\\ 
    &= W P_{\text{signal}} + P_{\text{non-path}}.
\end{align}
where $P_{\text{system,added}} = P_{\text{signal}} -  P_{\text{source,out}}$ is the power added (e.g., contributed) to the signal output power solely by the device or cascade. This leads to the practical application of \textit{W} in determining the wasted power $P_{\text{non-signal}}$ across a cascaded system referring to either the signal output power or input power, although it is most sensible to relate $W$ to the signal output power \cite{Rappaport2024Microwave, Rappaport2024Access},
\vspace{-3pt}
\begin{equation}
\label{powerwasted}
P_{\text{non-signal}} =  (W-1) P_{\text{signal}} = (W-1) P_{\text{source,out}}\prod_{i = 1}^{N}G_{i},
\end{equation}
where $G_{i}$ denotes the gain of the \(i^{th}\) stage in the cascade, and $N$ is the total number of cascaded components. Eqs. (\ref{defW}), (\ref{powercontotal}), and (\ref{powerwasted}) provide methods to analyze, design, and evaluate the power efficiency of circuits and systems. Notably, $W = 1$ indicates no wasted power, while $W = \infty$ implies that all power is wasted and no power is delivered to the output. Like $F$, the parameter $W$ is independent of the signal powers. Following eq. (\ref{powerwasted}), we observe $F-1$ in \eqref{addnoise} represents the added noise power relative to the input noise, and analogously, $W-1$ in \eqref{powerwasted} represents the total wasted power relative to the output signal power.

$W$ for a cascade of \textit{N} signal-path components can be computed from \eqref{defW} and \eqref{powercontotal} to be given by \eqref{Wcas} where device \textit{N} is at the sink of the cascade \cite{Rappaport2024Microwave,kanhere2022power,ying2023waste, Murdock2011GC, murdock2014JSAC}:
\begin{equation}
W = W_{N} + \frac{(W_{N-1} - 1)}{G_{N}} + \cdots + \frac{(W_{1} - 1)}{\prod_{i = 2}^{N} G_{i}}.
\label{Wcas}
\end{equation}

Eqn. \eqref{Wcas} resembles the cascaded Noise Factor in \eqref{Fcas}.

\section{W Analysis for Parallel Communication Systems}

MISO, SIMO, and MIMO are widely used architectures in wireless networks to improve system performance, capacity, and reliability. However, power efficiency analysis of these paralleled systems using $W$ has not been explored previously. 

\subsection{W for Multiple-Input Single-Output (MISO) System}
\label{sec:MISO}

To study MISO, we consider two transmitters (TXs) and one receiver (RX). Each transmitter communicates with the receiver over uncorrelated channels with zero-mean additive white Gaussian noise (AWGN).

\subsubsection{W for non-coherent combining MISO}

\begin{figure}[!t]
    \centering
    \includegraphics[width=2.8in]{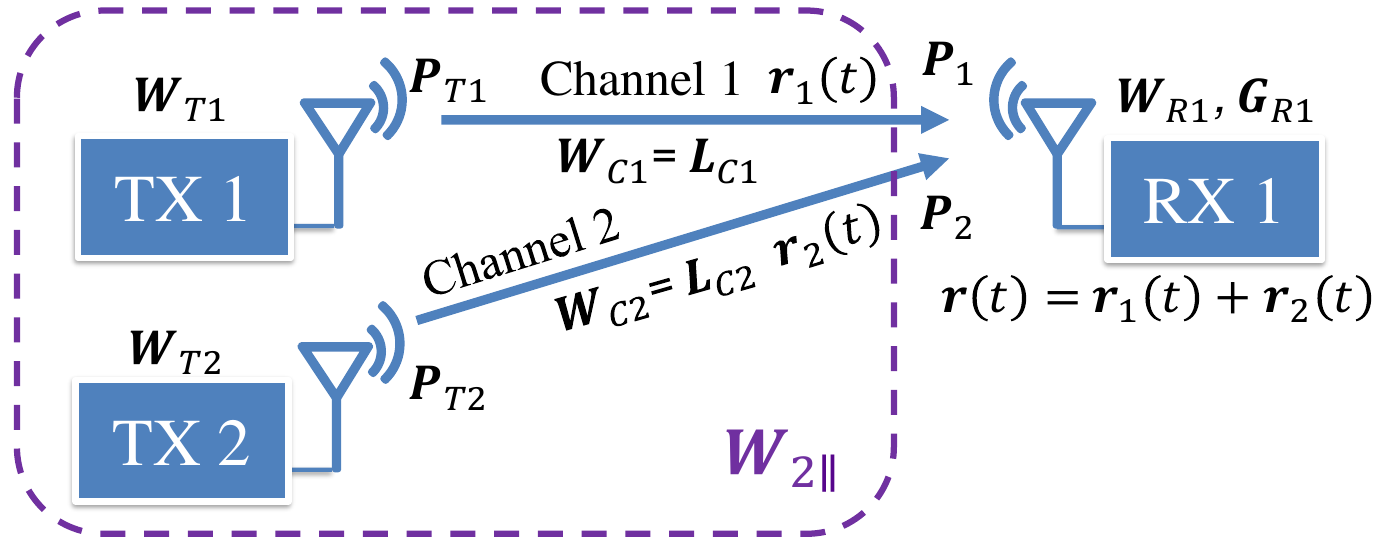}
    \caption{A two TX and one RX MISO communication system.}
    \label{figp1}
    \vspace{-8pt}
\end{figure}

In Fig. \ref{figp1}, TX1 has a Waste Factor $W_{T1}$ with channel having $W_{C1} = L_{C1}$ and received power $P_{1}$ into RX1 with $W_{R1}$ and receiver gain $G_{R1}$. Similarly, TX2 has its own channel and provides a received signal power $P_{2}$ to RX1. Using results of a lossy channel from \cite{Rappaport2024Microwave}, $W$ for each of these two parallel systems, when assuming non-coherent combining, allow the received powers to be combined \cite{Shu2014Beamcomb} to yield a single FoM $W_{\text{2}\mathbin{\parallel}}$. For two received signals \( r_1(t) \) and \( r_2(t) \) with received powers \( P_1 \) and \( P_2 \), respectively, non-coherent combining has no phase information. The combined RX power \( P_\text{signal,noncoh} \) in non-coherent combining at the RX antenna is typically the sum of individual powers:\(P_\text{signal,noncoh} = P_1 + P_2\) \cite{Shu2014Beamcomb}. The non-coherent combining assumes that the phases of the incoming signals from each TX are uniformly and identically distributed (i.i.d.), allowing for the straightforward addition of the powers~\cite{rappaport2024wireless}.
Using \eqref{defW} and \eqref{Wcas}, we find for a MISO system:
\begin{align}
    \label{p1}
    W_{\text{2}\mathbin{\parallel}}^\text{noncoh} 
    &= \frac{P_{\text{consumed,path}}}{P_\text{signal,noncoh}}  \nonumber\\
    &= \frac{{P_{1} \left(W_{C1} +\frac{W_{T1}-1}{G_{C1}} \right)}+ {P_{2} \left(W_{C2} +\frac{W_{T2}-1}{G_{C2}} \right)}}{P_{1} + P_{2}}.
\end{align}

Assuming the power received from each TX is proportional to some power \(P\) with coefficients \(\gamma_1\) and \(\gamma_2\) for the first and second transmitters, respectively, such that \([P_1, P_2] = P[\gamma_1, \gamma_2]\), eq. \eqref{p1} simplifies accordingly:
\begin{equation}
    W_{\text{2}\mathbin{\parallel}}^\text{noncoh} = \frac{{\gamma_1 \left(W_{C1} +\frac{W_{T1}-1}{G_{C1}} \right)}+ {\gamma_2 \left(W_{C2} +\frac{W_{T2}-1}{G_{C2}} \right)}}{\gamma_1 + \gamma_2}
    \label{W2par}
\end{equation}

\begin{figure}[!t]
    \centering
    \includegraphics[width=2.8in]{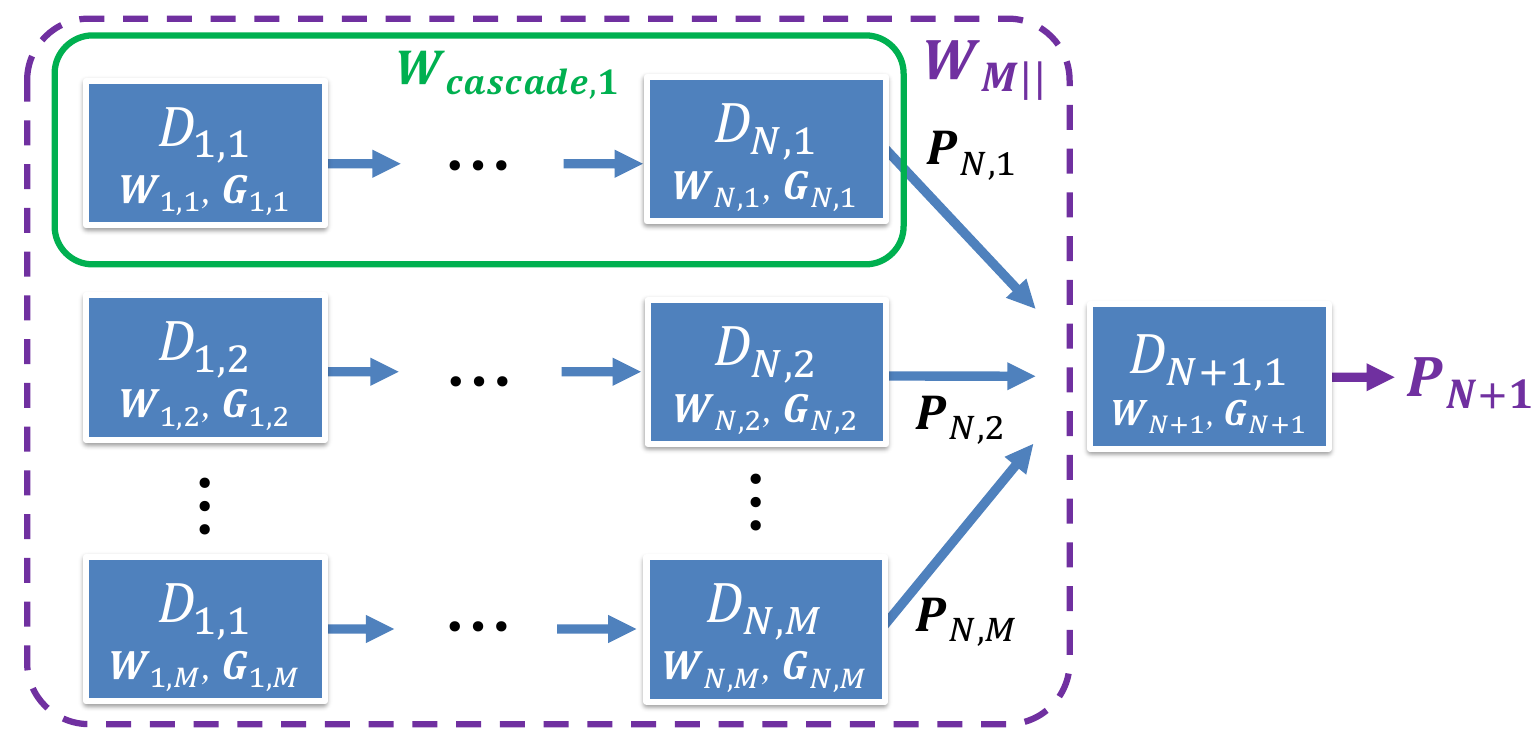}
    \caption{A MISO system with M paralleled input cascade.}
    \label{fig:MISO_M}
    \vspace{-15pt}
\end{figure}

In Fig. \ref{fig:MISO_M}, we envisage M parallel cascades, each cascade comprising \(N\) devices, with the output signal power combined at device \(N+1\) using either coherent or non-coherent combining. To ascertain $W$ for the entire paralleled system with a single output node, we first compute the \(W\) for cascade \(m\) using eq. \eqref{Wcas}:
\begin{equation}\label{p5}
    W_\text{cascade,m} = W_{N,m} + \frac{(W_{N-1,m}-1)}{G_{N,m}} + \cdots + \frac{(W_{1,m}-1)}{\prod_{i=2}^N G_{i,m}}.
\end{equation}

Subsequently, \(W\) for the entire paralleled system using non-coherent combining at ${N+1}^{th}$ device is given by:
\begin{equation}
\label{W_par_inco}
    W_{\text{M}\mathbin{\parallel}}^\text{noncoh} = \frac{\sum_{i=1}^{\text{M}} \left({P_{N,i} \times W_\text{cascade,i}}\right)}{\sum_{i=1}^{\text{M}} {P_{N,i}}}.
\end{equation}

Here, we assume the power output coming out of each cascade is represented by $[P_{N,1}, P_{N,2},\ldots, P_{N,M}] = P[\gamma_1,  \gamma_2,\ldots,\gamma_M]$, where \(P\) is a constant base power and \(\gamma_i\) are scaling factors that illustrate the relative contribution to the total received power from each cascade output. Then \eqref{W_par_inco} can be denoted as:
\begin{equation}
\label{W_par_inco_gamma}
    W_{\text{M}\mathbin{\parallel}}^\text{noncoh}  = \frac{\sum_{i=1}^{\text{M}} {\gamma_{i} W_\text{cascade,i}}}{ \sum_{i=1}^{\text{M}} {\gamma_{i}} }.
\end{equation}

\subsubsection{W for coherent combining MISO} 

For coherent combining, the receiver must be phase-synchronized with signals \( r_1(t) \) and \( r_2(t) \). If \( r_1(t) \) and \( r_2(t) \) are phase-aligned and have powers \( P_1 \) and \( P_2 \), the combined signal power at the RX antenna \( P_\text{signal,coh} \) is given by the square of the magnitude of the vector sum of the two signals \cite{Shu2014Beamcomb}:
\begin{equation}
     P_\text{signal,coh} = \left|\sqrt{P_1} + \sqrt{P_2}\right|^2,
     \label{p3}
\end{equation}
where \( \theta_1 \) and \( \theta_2 \) are the phases of \( r_1(t) \) and \( r_2(t) \) respectively.

Considering coherent combining in a M TX single RX MISO system, we assume the power output of \(i^{th}\) TX  as \(P_{Ti}\), and the power received from transmitter \(i\) at the RX before combining as $P_{i}$, and the power into the RX after coherent combining at the RX antenna as $P_\text{signal,coh}$. Here we assume  $[P_{1}, P_{2},\ldots,P_{M}] = P[\gamma_1,  \gamma_2,\ldots,\gamma_M]$, where we show that the received power from each TX has a relationship with each other, and all are related to the ratio of some $P$.

Using eq. \eqref{defW} and \eqref{p3}, and following eq. \eqref{W2par}, \textit{W} for this coherent combining MISO can be reformulated as:
\begin{equation}
\label{p4}
    W_{\text{M}\mathbin{\parallel}}^\text{coh} = \frac{\sum_{i=1}^{\text{M}} \left[ {\gamma_{i} \left(W_{Ci} +\frac{W_{Ti}-1}{G_{Ci}} \right)} \right]}{ \left|\sum_{i=1}^{\text{M}} {\sqrt{\gamma_i} } \right|^2 }.
\end{equation}

Based on \eqref{p4}, considering a general MISO system in Fig. \ref{fig:MISO_M} with a single output, \textit{W} for the entire parallel system using coherent combining is:
\begin{equation}
\label{W_par_co}
    W_{\text{M}\mathbin{\parallel}}^\text{coh} = \frac{\sum_{i=1}^{\text{M}} \left({P_{N,i} \times W_\text{cascade,i}}\right)}{\left|\sum_{i=1}^{\text{M}} {\sqrt{P_{N,i}} } \right|^2 } 
    = \frac{\sum_{i=1}^{\text{M}} \left( {\gamma_{i} W_\text{cascade,i}} \right)}{ \left|\sum_{i=1}^{\text{M}} {\sqrt{\gamma_i} } \right|^2 }.
\end{equation}

We see for MISO in general, using $W_{\text{M}\mathbin{\parallel}}$ for either non-coherent \eqref{W_par_inco_gamma} or coherent \eqref{W_par_co} combining, we can calculate \textit{W} of a MISO system shown in Fig. \ref{fig:MISO_M} as:
\begin{equation}
\label{WMISO}
    W_{\text{MISO}} =  W_{N+1}  + \frac{(W_{\text{M}\mathbin{\parallel}} - 1)}{G_{N+1}}.
\end{equation}
where $W_{N+1}$ and $G_{N+1}$ represents the Waste Factor and gain of the ${(N+1)}^{th}$ single output device.

\subsection{W for Single-Input Multiple-Output (SIMO) System}
\label{sec:SIMO}
\begin{figure}[!t]
    \centering
    \includegraphics[width=2.8in]{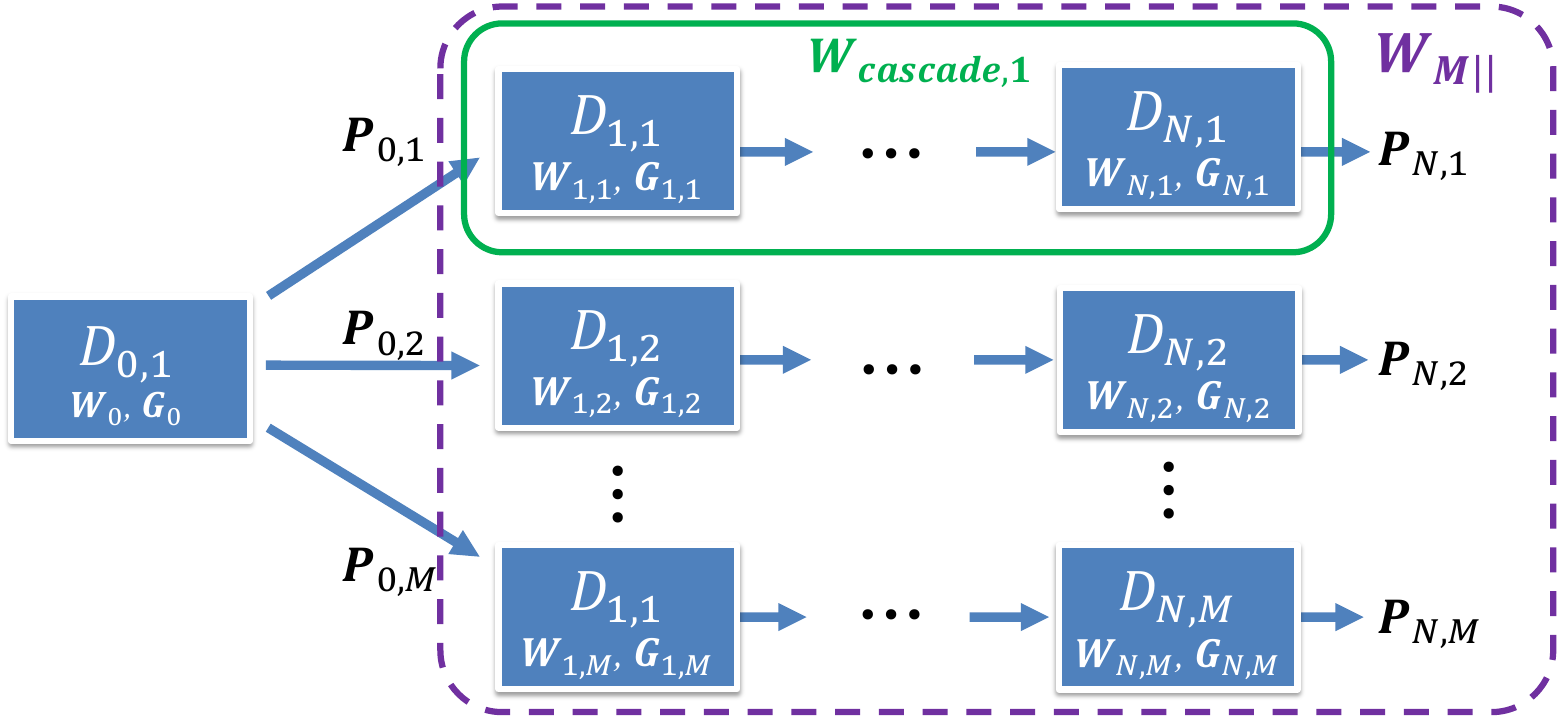}
    \vspace{-5pt}
    \caption{A SIMO system with M paralleled output cascade.}
    \label{fig:SIMO}
    \vspace{-15pt}
\end{figure}

Consider a SIMO system as depicted in Fig. \ref{fig:SIMO}. Waste Factor for the system within the dashed box is analogous to the MISO structure. Here, we need to consider coherent combining or non-coherent combining for the output of the parallel cascade in the dashed box of Fig. \ref{fig:SIMO}. Waste factor for a SIMO system is derived using a cascade comprising two main components. The first component denoted as \(D_{0,1}\), is characterized by a Waste Factor \(W_0\) and a gain \(G_0\). The second component includes the devices within the dashed box of Fig. \ref{fig:SIMO}, and here we need the gain and Waste Factor of the dashed box.

\subsubsection{W for non-coherent combining SIMO} Consider non-coherent combining at the output of paralleled cascades, and the input power of each cascade is independent. Assume $[P_{N,1}, P_{N,2},\ldots, P_{N,M}] = P[\gamma_1, \gamma_2,\ldots,\gamma_M]$, then Waste Factor for the dashed box using non-coherent combining at its output is:
{\small  \begin{equation}
\label{SIMO_noncoh}
        W_{\text{M}\mathbin{\parallel}}^\text{noncoh} =
        \frac{\sum_{i=1}^{\text{M}} \left({P_{N,i} \times W_\text{cascade,i}}\right)}{\sum_{i=1}^{\text{M}} {P_{N,i}}} 
        = \frac{\sum_{i=1}^{\text{M}} \left({\gamma_{i} \times W_\text{cascade,i}}\right)}{\sum_{i=1}^{\text{M}} {\gamma_{i}}},  
\end{equation}}
{\noindent  \noindent and \(G_{\text{M}\mathbin{\parallel}}^\text{noncoh} =\left(\sum_{i=1}^{\text{M}} {P_{N,i}}\right)/  \left(\sum_{i=1}^{\text{M}} {P_{0,i}}\right) \) is calculated as the ratio of the sum of the total output power of the devices in the dashed box using non-coherent combining to the sum of the input powers.}

\subsubsection{W for coherent combining SIMO} If we consider coherent combining at each output of the parallel cascade, and assume each cascade has an independent input power from device $D_{0,1}$, and $[P_{N,1}, P_{N,2},\ldots, P_{N,M}] = P[\gamma_1, \gamma_2,\ldots,\gamma_M]$. Then Waste Factor for the system in the dashed box is:
\begin{equation}
\label{SIMO_coh}
        W_{\text{M}\mathbin{\parallel}}^\text{coh} 
        = \frac{\sum_{i=1}^{\text{M}} \left({P_{N,i} \times W_\text{cascade,i}}\right)}{\left|\sum_{i=1}^{\text{M}} {\sqrt{P_{N,i}} } \right|^2 } 
        = \frac{\sum_{i=1}^{\text{M}} \left( {\gamma_{i} W_\text{cascade,i}} \right)}{ \left|\sum_{i=1}^{\text{M}} {\sqrt{\gamma_i} } \right|^2 },
\end{equation}
and \(G_{\text{M}\parallel}^{\text{coh}} = {\left(\sum_{i=1}^M \sqrt{P_{N,i}}\right)^2}/{\left(\sum_{i=1}^M P_{0,i}\right)}\) is the ratio of the total output power using coherent combining of the devices in the dashed box to the sum of the input powers. Based on eq. \eqref{SIMO_noncoh} and \eqref{SIMO_coh}, the overall Waste Factor for the general SIMO system can be expressed as:
\vspace{-6pt}
\begin{equation}
\label{W_SIMO}
        W_{\text{SIMO}} =  W_{\text{M}\mathbin{\parallel}}  + \frac{(W_0 - 1)}{G_{\text{M}\mathbin{\parallel}}}.
\end{equation}
where $G_{\text{M}\mathbin{\parallel}}$ is different depending on whether coherent or non-coherent combing at the output of parallel cascades.

\subsection{W for Multiple-Input Multiple-Output (MIMO) System}
\label{sec:MIMO}

\subsubsection{W for non-coherent combining MIMO}

\begin{figure}[!t]
    \centering
    \includegraphics[width=2.7in]{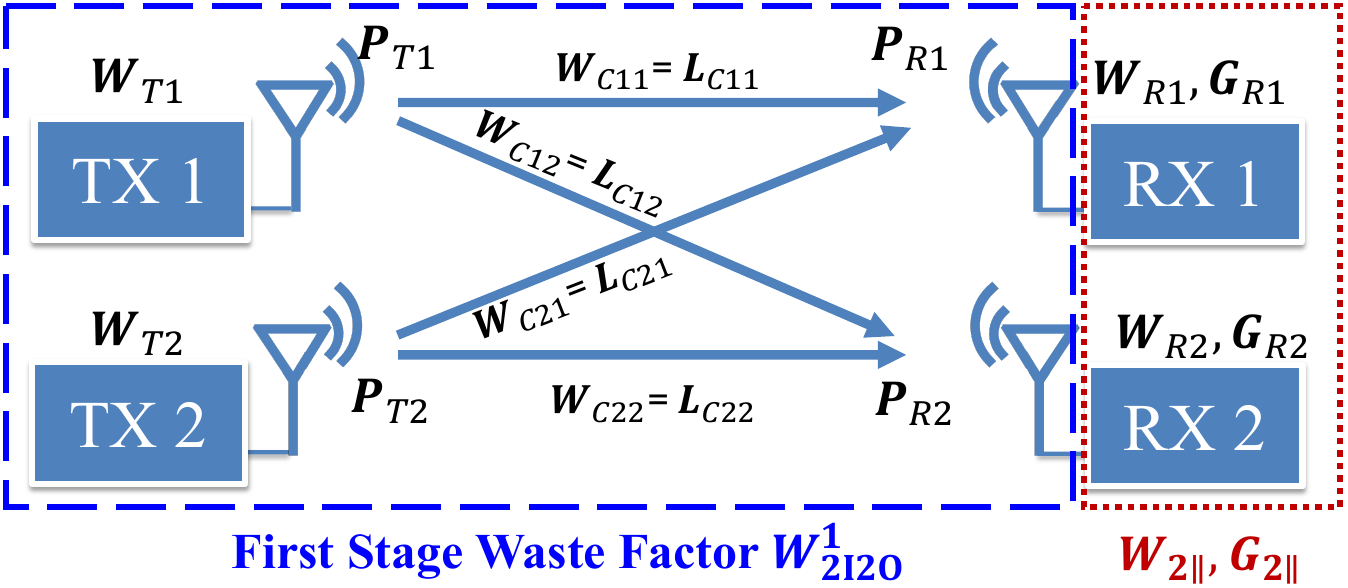}
    \caption{A two-input and two-output MIMO system.}
    \label{fig:MIMO}
    \vspace{-15pt}
\end{figure}

For a 2-Input 2-Output (2I2O) MIMO system in Fig. \ref{fig:MIMO}, the received power of each RX using non-coherent combining is: 
\begin{equation}
\label{PRX_noncoh}
\begin{bmatrix}
P_{R1}^\text{noncoh} & P_{R2}^\text{noncoh}
\end{bmatrix}
= \begin{bmatrix}
P_{T1} & P_{T2}
\end{bmatrix}
\begin{bmatrix}
W_{C11}^{-1} & W_{C12}^{-1} \\
W_{C21}^{-1} & W_{C22}^{-1}
\end{bmatrix}.
\end{equation}
where $P_{Ri}^\text{noncoh}$ is the non-coherently combined power at the antenna of RX $i$. The total signal-path power consumption of the system before the receiver using non-coherent combining is:
\vspace{-6pt}
\begin{equation}
P_{\text{consumed,path}}^\text{noncoh} = \sum_{i=1}^{2} { \left({P}_{Ri}^\text{noncoh} W_{2\parallel}^\text{noncoh} \right) },
\end{equation}
where $W_{2\parallel}^\text{noncoh}$ represents Waste Factor for a 2TX paralleled system together with the channel, which is the same as \eqref{W2par}. Before the power goes into the receivers, we define the first stage Waste Factor of the 2I2O system using non-coherent combining in the dashed box of Fig. \ref{fig:MIMO}:
\begin{equation}
\label{W2I2O_1st}
W_{\text{2I2O}}^{1,\text{noncoh}} = \frac{\sum_{i=1}^{2} { \left({P}_{Ri}^\text{noncoh} W_{2\parallel}^\text{noncoh} \right) }}{\sum_{i=1}^{2} { \left({P}_{Ri}^\text{noncoh} \right) }}.
\end{equation}

Letting $[P_{R1}^\text{noncoh},P_{R2}^\text{noncoh}]=P^\text{noncoh}[\gamma_1,\gamma_2]$, then \eqref{W2I2O_1st} can be written as:
\vspace{-3pt}
\begin{equation}
W_{\text{2I2O}}^{1,\text{noncoh}} = \frac{\sum_{i=1}^{2} { \left({\gamma}_{i} W_{2\parallel}^\text{noncoh} \right) }}{\sum_{i=1}^{2} { \left({\gamma}_{i} \right) }}.
\end{equation}

To encapsulate the entire power efficiency of the system, including the receivers, the complete Waste Factor for a 2I2O system using non-coherent combining ($W_{\text{2I2O}}^\text{noncoh}$) is calculated by cascading the first-stage $W_{\text{2I2O}}^{1,noncoh}$ and $W_{\text{2}\mathbin{\parallel}}^\text{noncoh}$ in the dotted box from the RX side based on \eqref{W_par_inco}:
\begin{equation}
\label{W2I2O_noncoh}
W_{\text{2I2O}}^\text{noncoh} = W_{\text{2}\mathbin{\parallel}}^\text{noncoh} + \frac{(W_{\text{2I2O}}^{1,\text{noncoh}} - 1)}{G_{\text{2}\mathbin{\parallel}}^\text{noncoh}}.
\end{equation}

\begin{figure*} [!t]
\centering
\includegraphics[width=14cm]{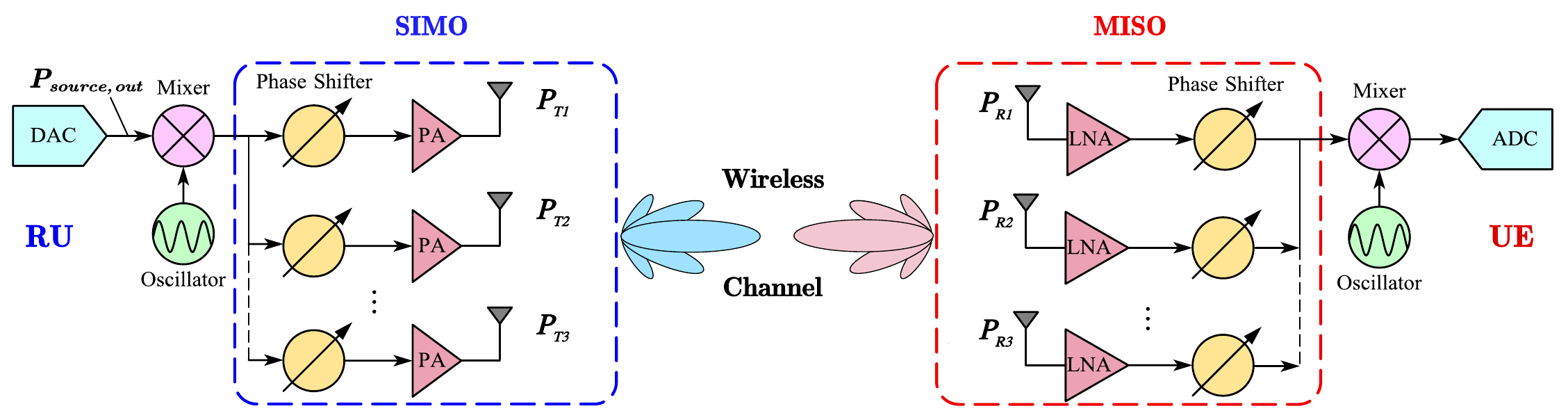}
\caption{General structure of the RU and UE.}
\label{fig:RANStructure}
\vspace{-15pt}
\end{figure*}

\subsubsection{W for coherent combining MIMO}

The received power of each RX using coherent combining is:
{\small
\begin{equation}
\label{PRX_coh}
\begin{bmatrix}
P_{R1}^\text{coh} & P_{R2}^\text{coh}
\end{bmatrix}
= \left(\begin{bmatrix}
\sqrt{P_{T1}} & \sqrt{P_{T2}}
\end{bmatrix}
\begin{bmatrix}
\sqrt{W_{C11}^{-1}} & \sqrt{W_{C12}^{-1}} \\
\sqrt{W_{C21}^{-1}} & \sqrt{W_{C22}^{-1}}
\end{bmatrix}\right)^{2},
\end{equation}
}
where the exponent \(^2\) denotes an element-wise square of each term in the resulting matrix.

The total signal-path power consumption of the system before the coherently combined received  power going into RX is given by:
\vspace{-6pt}
\begin{equation}
P_{\text{consumed,path}}^\text{coh} = \sum_{i=1}^{2} { \left({P}_{Ri}^\text{coh} W_{2\parallel}^\text{coh} \right) },
\end{equation}
where $P_{Ri}^\text{coh}$ is the coherently combined power at the antenna of RX $i$, and $W_{2\parallel}^\text{coh}$ represents Waste Factor for a 2-TX paralleled system together with the channel, which is the same as \eqref{p4} with M = 2. Before the power goes into the receivers, we define the first-stage Waste Factor of the 2I2O system using coherent combining as:
\vspace{-6pt}
\begin{equation}
W_{\text{2I2O}}^{1,\text{coh}} = \frac{\sum_{i=1}^{2} { \left({P}_{Ri}^\text{coh} W_{2\parallel}^\text{coh} \right) }}{\sum_{i=1}^{2} { \left({P}_{Ri}^\text{coh} \right) }}.
\label{W2I2O_1st_coh}
\end{equation}
Letting $[P_{R1}^\text{coh},P_{R2}^\text{coh}]=P^\text{coh}[\gamma_1,\gamma_2]$, then \eqref{W2I2O_1st_coh} can be written as:
\begin{equation}
W_{\text{2I2O}}^{1,\text{coh}} = \frac{\sum_{i=1}^{2} { \left({\gamma}_{i} W_{2\parallel}^\text{coh} \right) }}{\sum_{i=1}^{2} { \left({\gamma}_{i} \right) }}.
\label{W2I2O_1st_coh_gamma}
\end{equation}

To encapsulate the entire power efficiency of the system, including the receivers, the complete Waste Factor for a 2I2O system using coherent combining ($W_{\text{2I2O}}^\text{coh}$) is calculated by cascading the first stage $W_{\text{2I2O}}^{1,\text{coh}}$ and $W_{\text{2}\mathbin{\parallel}}^\text{coh}$ from receiver side based on \eqref{W_par_co}:
\begin{equation}
W_{\text{2I2O}}^\text{coh} = W_{\text{2}\mathbin{\parallel}}^\text{coh} + \frac{(W_{\text{2I2O}}^{1,\text{coh}} - 1)}{G_{\text{2}\mathbin{\parallel}}^\text{coh}}.
\label{W2I2O_coh}
\end{equation}

\subsubsection{W for general MIMO system}
If we assume proportional combined powers of each RX, which means that the power after coherent or non-coherent combining, the received power of each RX has a relationship with each other, and all are related to the ratio of some $P$,
\begin{equation} 
[P_{R1},P_{R2},\ldots,P_{R\text{M}} ] = P[\gamma_1,\gamma_2,\ldots,\gamma_{\text{M}} ],
\end{equation}
then \eqref{W2I2O_1st} and \eqref{W2I2O_1st_coh_gamma} can be extended to:
\begin{equation}
W_{\text{MIMO}}^{1} = \frac{\sum_{i=1}^{\text{M}} { \left({\gamma}_{i} W_{\text{M}\parallel} \right) }}{\sum_{i=1}^{\text{M}} { \left({\gamma}_{i} \right) }}.
\label{WMIMO_1st}
\end{equation}

Eventually, $W$ for the generalized MIMO system is:
\begin{equation}
W_{\text{MIMO}} =  W_{\text{M}\mathbin{\parallel}}  + \frac{(W_{\text{MIMO}}^{1} - 1)}{G_{\text{M}\mathbin{\parallel}}}.
\end{equation}
where $G_{\text{M}\mathbin{\parallel}}$ is the gain of the paralleled RXs, which can be calculated based on the ratio of the total output power of RXs to the input power of RXs:
\vspace{-6pt}
\begin{equation}
G_{\text{M}\mathbin{\parallel}} = \frac{ \sum_{i=1}^{\text{M}} P_{Ri} G_{Ri}}{\sum_{i=1}^{\text{M}} P_{Ri}}.
\end{equation}
\vspace{-6pt}

\section{Waste Factor Analysis in Communication Systems}
\label{sec:RAN}
Here, we discuss the calculation of the \(W\) for a RAN depicted in Fig. \ref{fig:RANStructure}. Waste Factor, \(W_{C} = L_{C}\), is crucial for assessing the energy efficiency of communication channels to include mobile users~\cite{Rappaport2024Microwave, murdock2014JSAC, Rappaport2024WCNCTutorial, Rappaport2024Access}. When considering links with antenna gain, we define an effective channel Waste Factor $W_{C}^{\text{eff}} = {L_{C}}/\left({G_{\text{TX}}^{\text{ant}} G_{\text{RX}}^{\text{ant}}}\right) = 1/G_{C}^{\text{eff}}$ to simplify the overall channel loss, where \(G_{\text{TX}}^{\text{ant}}\) and \(G_{\text{RX}}^{\text{ant}}\) represent the antenna gains at the transmitter and receiver, respectively.

\subsection{Waste Factor Evaluation in RAN Site Chain}

We first study a single TX at the base station (BS), specifically the radio unit (RU). This RU is composed of a digital-to-analog converter (DAC), a mixer, M phase shifters (PSs), M power amplifiers (PAs), and M antennas, with the power output from the DAC denoted as $P_{\text{source,out}}$. This configuration is analyzed as a SIMO system, where each parallel cascade in the dashed box includes a PS, a PA, and an antenna.

Let us assume non-coherent combining at the receiver, which implies that the incoming signals are combined based on their power levels without considering their phase information. Refer to Section~\ref{sec:SIMO}, where $W$ for the system is calculated based on the performance of individual components within each parallel cascade.\vspace{-6pt}
\begin{equation}
\label{W_par_RU}
    W_{\text{M}\mathbin\parallel}^{\text{RU}} = \frac{\sum_{i=1}^{M} {P_{T,i} \times  \left( W_{\text{Ant}} + \frac{W_{\text{PA}} - 1}{G_{\text{Ant}}} + \frac{W_{\text{PS}} - 1}{G_{\text{PA}} G_{\text{Ant}}}   \right)}}{\sum_{i=1}^{M} {P_{T,i}}},
\end{equation}
where $P_{T,i}$ represents the transmit power from the $i^{th}$ antenna element. Next, we determine \textit{W} for the entire RU, which includes all parallel cascades and the mixer, denoted as $W_{\text{RU}}$, incorporating the overall gain of the paralleled structure in the RU:
\begin{equation}
\label{W_RU}
        W_{\text{RU}} =  W_{\text{M}\mathbin\parallel}^{\text{RU}}  + \frac{(W_{\text{Mix}} - 1)}{G_{\text{M}\mathbin\parallel}^{\text{RU}}},
\end{equation}
where $ G_{\text{M}\mathbin\parallel} = {\sum_{i=1}^{M} {P_{T,i}}}/{\left(P_{\text{source,out}}G_{\text{Mix}}\right)}$ is the gain of the parallel cascade in the dashed box of RU.

For a system with an identical paralleled cascade, the transmit power from each antenna element is the same, and \textit{W} in \eqref{W_RU} can be simplified to:
\begin{equation}
\label{W_RU_simple}
W_{\text{RU}} = W_{\text{Ant}} + \frac{W_{\text{PA}} - 1}{G_{\text{Ant}}} + \frac{W_{\text{PS}} - 1}{G_{\text{PA}} G_{\text{Ant}}} + \frac{W_{\text{Mix}} - 1}{G_{\text{PS}} G_{\text{PA}} G_{\text{Ant}}}.
\end{equation}

Extending the analysis to the receiver side alone, the user equipment (UE) can be treated as a MISO system, and $W_{\text{UE}}$ is calculated assuming the same components across all parallel cascades in a single UE using non-coherent combining, based on eq. \eqref{W_par_inco_gamma} for non-coherent combing MISO in Section~\ref{sec:MISO}:
\begin{equation}
\label{WUE}
W_{\text{UE}} = W_{\text{Mix}} + \frac{W_{\text{PS}} - 1}{G_{\text{Mix}}} 
+ \frac{W_{\text{LNA}} - 1}{G_{\text{Mix}} G_{\text{PS}}} + \frac{W_{\text{Ant}} - 1}{G_{\text{Mix}} G_{\text{PS}} G_{\text{LNA}}}
\end{equation}

Finally, Waste Factor for the entire RAN cascade, including a RU, a wireless channel, and a UE is derived using eq. \eqref{Wcas}:
\begin{equation}
    \label{eq:overall_W}
    W_{\text{RAN}} = W_{\text{UE}}+\frac{W_{C}^{\text{eff}}-1}{G_{\text{UE}}} + \frac{W_{\text{RU}}-1}{G_{C}^{\text{eff}}G_{\text{UE}}}.
\end{equation}

\section{Simulation and Results Discussion}

Simulations were conducted within a 28 GHz Coordinated Multi-Point (CoMP) communication system to assess the effectiveness of using \textit{WF} for a MIMO system. The network featured 512 UEs, with the number of BSs varying from 1 to 20, the simulation cell has a radius of 1 km, and each BS encompasses a radius of 200 m. It is assumed that only users within these ranges are served. Refer to Section \ref{sec:RAN}, BSs and UEs using non-coherent combining were positioned at heights of 15 and 1.5 meters, respectively, with a separation of 300 m between the nearest BSs. Antenna gains for BSs and UEs were set to 26 dB and 6 dB, respectively, and the bandwidth is 400 MHz \cite{shakya2022ICCPara}. This configuration represents a typical urban microcell (UMi) environment, aiming for an SNR of 10 dB at the UEs. The transmit power was dynamically adjusted to meet the SNR requirement at each UE, with a cap of 100 Watts for each BS. The path loss of the wireless channel, incorporating a line-of-sight (LOS) path loss exponent of 2.27 and a shadow fading standard deviation of 8.15 dB, was derived from urban propagation research conducted by NYU WIRELESS \cite{shakya2024radio}. Additionally, UEs situated within the coverage areas of multiple BSs received combined power from all such BSs, demonstrating the collaborative transmission feature of the CoMP system \cite{MacCartney2019TWC}.

\begin{figure}[!t]
    \centering
    \begin{overpic}[width=3.1in]{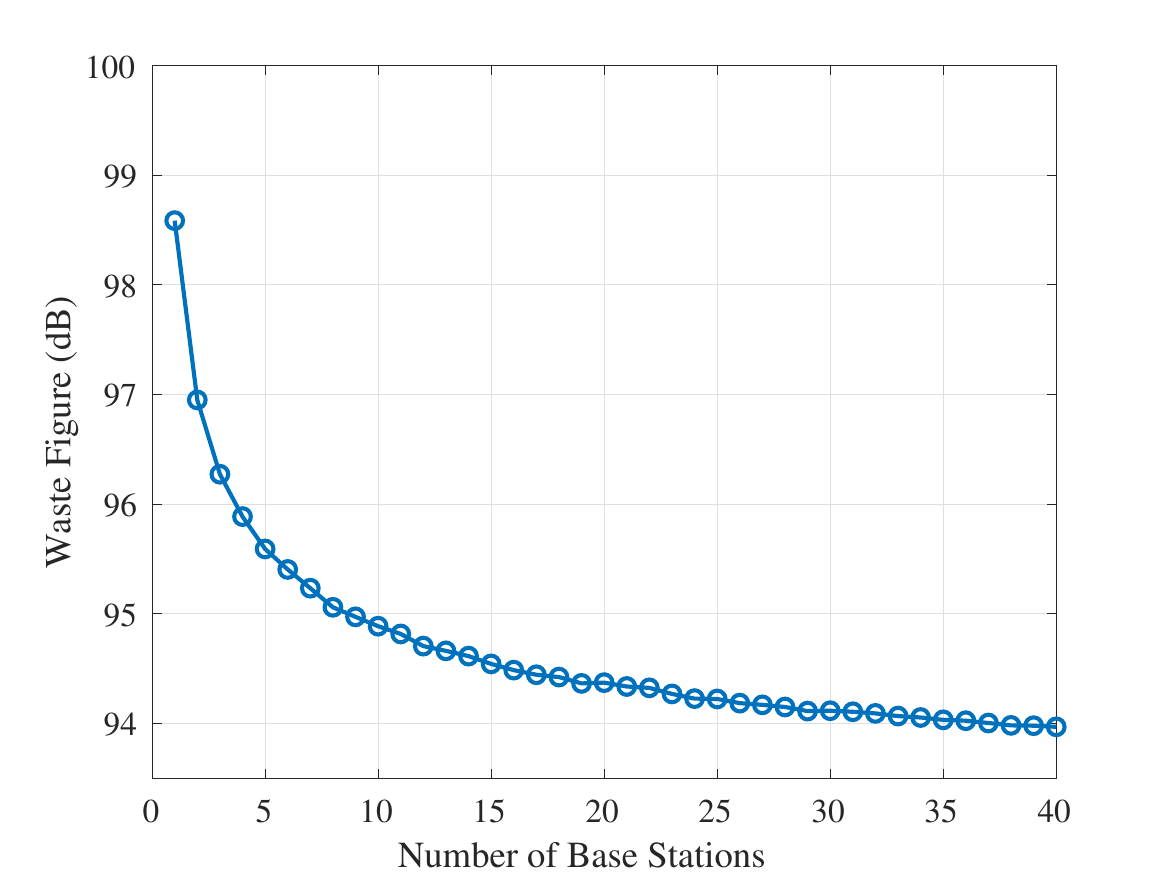}
        \put(35,22){\includegraphics[width=1.6in]{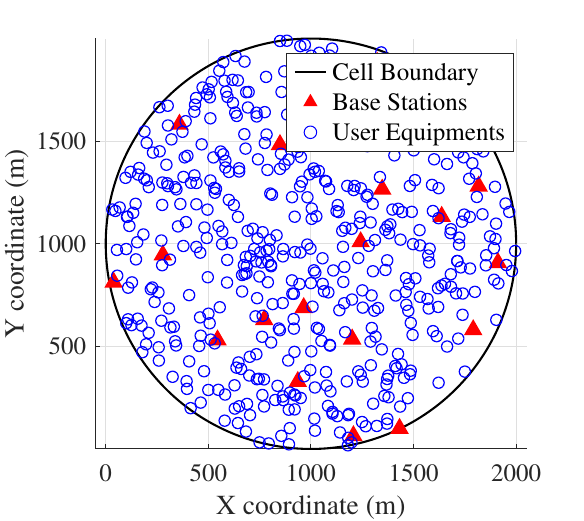}} % Adjust the positioning and size here
    \end{overpic}
    \vspace{-10pt}
    \caption{Waste Figure with different base station numbers in a 28 GHz coordinated multi-point communication network.}
    \label{fig:WF_28GHz}
    \vspace{-15pt}
\end{figure}

Simulation results, shown in Fig. \ref{fig:WF_28GHz}, demonstrated that as the number of BSs increased, the \textit{WF} decreased. This indicates that network densification improves the overall system power efficiency. Additionally, when the number of BSs is small, the slope of Fig. \ref{fig:WF_28GHz} is steeper, indicating that \textit{WF} drops (improves) significantly in a CoMP scenario with increasing BS density, \textit{WF} drops significantly as the number of BSs increases. However, as the number of BSs continues to increase, the \textit{WF} reaches a lower limit. The results suggest that a combination of network densification and strategic component optimization can lead to better system design.

\section{Conclusions}

This paper introduces the Waste Factor theory for complex paralleled architectures like MISO, SIMO, and MIMO. The work compares \textit{W} and \textit{WF} with conventional energy efficiency metrics, highlighting advantages in providing a comprehensive and flexible approach to power efficiency analysis. By extending the analysis of power waste to include the impact of communication channels, RAN chains, and paralleled systems, this work offers a complete framework for assessing the power efficiency of modern wireless networks. Future directions include applying \textit{W} to evaluate power consumption in various beamforming structures and RF chain on/off strategies, as well as exploring ring structure network energy consumption.

\bibliographystyle{IEEEtran}
\bibliography{reference}

\end{document}